\documentclass[a4paper]{jpconf}
\usepackage{graphicx}
\usepackage{amsmath,bm}
\usepackage{verbatim}
\usepackage{amsthm,mathtools}

\newcommand{\bra}[1]{\left\langle{#1}\right|}
\newcommand{\ket}[1]{\left|{#1}\right\rangle}

\newcommand{\be}{\begin{equation}}
\newcommand{\ee}{\end{equation}}

\newcommand{\A}{{\cal A}}
\newcommand{\ssigma}{{\bm \sigma}}

\newcommand{\Sb}{{\bm S}}

\begin{document}
\title{Merits of using density matrices instead of wave functions in the stationary Schr\"odinger equation for systems with symmetries
}

\author{E Shpagina$^{1,2,3}$, F Uskov$^1$, N Il'in$^{1}$, O Lychkovskiy$^{1,4}$}

\address{$^1$ Skolkovo Institute of Science and Technology, Bolshoy Boulevard 30, bld. 1, Moscow 121205, Russia}
\address{$^2$ Bauman Moscow State Technical University, 2nd Baumanskaya str., 5, Moscow 105005, Russia}
\address{$^3$ NRU Higher School of Economics, Myasnitskaya 20, Moscow 101000, Russia}
\address{$^4$ Department of Mathematical Methods for Quantum Technologies, Steklov Mathematical Institute of Russian Academy of Sciences, Gubkina str., 8, Moscow 119991, Russia}

\ead{LeShpagina@yandex.ru }

\begin{abstract}
The stationary Schr\"odinger equation  can be cast in the form $H \rho = E \rho$, where $H$ is the system's Hamiltonian and $\rho$ is the system's density matrix. We explore the  merits of this form of the stationary Schr\"odinger equation, which we refer to as~SSE$_\rho$, applied to many-body systems with symmetries. For a nondegenerate energy level, the solution $\rho$ of the SSE$_\rho$ is merely a projection on the corresponding eigenvector. However, in the case of degeneracy $\rho$ is non-unique and not necessarily pure. In fact, it can be an arbitrary mixture of the degenerate pure eigenstates. Importantly, $\rho$ can always be chosen to respect all symmetries of the Hamiltonian, even if each pure eigenstate in the corresponding degenerate multiplet spontaneously breaks the symmetries. This and other features of the solutions of the SSE$_\rho$ can prove helpful by easing the notations and providing an unobscured insight into the structure of the eigenstates. We work out the SSE$_\rho$ for a general system of spins $1/2$ with Heisenberg interactions, and address  simple systems of spins $1$. Eigenvalue problem for quantum observables other than Hamiltonian can also be formulated in terms of density matrices. As an illustration, we provide an analytical solution to the eigenproblem ${\bf S}^2 \rho=S(S+1) \rho$, where $\bf S$ is the total spin of $N$ spins $1/2$, and $\rho$ is chosen to be invariant under permutations of spins. This way we find an explicit form of projections to the invariant subspaces of ${\bf S}^2$.
\end{abstract}

\section{General properties of the Stationary Schr\"odinger equation for density matrices}
The conventional form of the stationary Schr\"odinger equation (which we refer to as SSE$_\Psi$) reads
\begin{equation}\label{SSE}
 H \ket{\Psi} = E \ket{\Psi},
\end{equation}
where $H$ is the Hamiltonian of a quantum system, $E$ is its eigenenergy and $\ket{\Psi}$ is the corresponding eigenvector. Obviously, this equation implies an operator identity $ H \ket{\Psi}\bra{\Psi} = E \ket{\Psi}\bra{\Psi}$, where $\ket{\Psi}\bra{\Psi}$ is the projection onto the eigenvector $\ket{\Psi}$. One can extend this identity by considering an arbitrary density matrix $\rho$ instead of a projection operator. This way one obtains the stationary Schr\"odinger equation for density matrices:
\begin{equation}\label{SSErho}
H \rho = E \rho.
\end{equation}
In the present paper we explore the properties and the merits of this equation which we refer to as SSE$_\rho$. Our studies are somewhat close in spirit to the research avenue on the contracted Schr\"odinger equation, see e.g. \cite{herbert2002extensivity} and references therein. Some important differences will be discussed below when we apply the SSE$_\rho$ to a particular spin system in Section 2.

We remind that a density matrix should satisfies three conditions,
\be
\rho^\dagger=\rho,~~~~\tr \rho=1,~~~~\rho>0. \label{rho conditions}
\ee
The following relations between the SSE$_\Psi$ and  SSE$_\rho$ follow immediately.
\begin{enumerate}
\item Eqs. \eqref{SSE} and \eqref{SSErho} share the same set of eigenvalues $E$.
\item If a given eigenvalue $E$ is nondegenerate, then the corresponding $\ket{\Psi}$ and $\rho$ are related according to $\rho=\ket{\Psi}\bra{\Psi}$.
\item In the case of degeneracy any solution of the SSE$_\rho$ reads
\begin{align}\label{general solution}
 \rho = \sum_{i}^{} p_i \ket{\Psi_i} \bra{\Psi_i},~~~p_i\geq0, ~~~\sum_i p_i=1,
\end{align}
where vectors $\ket{\Psi_i}$ constitute a basis in the corresponding degenerate subspace of the Hamiltonian.
\end{enumerate}

Properties (ii) implies that in the nondegenerate case SSE$_\Psi$ and  SSE$_\rho$ are, in fact, identical up to notations (however, even in this case  SSE$_\rho$ can be more convenient compared to SSE$_\Psi$, in particular for spin systems, see e.g. \cite{lychkovskiy2017time}). An important advantage of the  SSE$_\rho$ shows up in the case of a degeneracy induced by some symmetry of the Hamiltonian. Assume that the Hamiltonian is symmetric under some symmetry group $G$, i.e.
\be
U\,H\,U^\dagger = H~~~~~\forall ~~U\in G,
\ee
where $U$ is a unitary operator. In this case the eigenbasis of $H$ is split into blocks which determine degenerate subspaces  invariant under the group $G$.  As a rule, a {\it spontaneous symmetry breaking} phenomenon occurs in some of this subspaces, which means that any eigen basis in such a subspace contains eigenvectors not invariant with respect to $G$. The most  trivial example of a system with the spontaneous symmetry breaking is a single spin $1/2$ with a vanishing Hamiltonian, $H=0$. Such Hamiltonian is invariant under  the group of rotations, however any its eigenstate (i.e. any pure state of a single spin $1/2$ ) lacks this symmetry. In general, an important class of spin systems with Heisenberg interactions demonstrate spontaneous symmetry breaking (either in the ground state or in excited states). Some examples of such systems will be considered in what follows. The phenomenon of the spontaneous symmetry breaking, while being of paramount importance for physics \cite{strocchi2008symmetry}, can sometimes cause various inconveniences. In particular, it obscures calculations of the correlation functions invariant with respect to $G$. In contrast to SSE$_\Psi$, one can always avoid the spontaneous symmetry breaking in the solutions of SSE$_\rho$, according to the following simple

\smallskip

\noindent
{\it Lemma.} Consider a Hamiltonian $H$ invariant under the symmetry group $G$ ($G$-invariant, for short). For any eigenvalue $E$ of this Hamiltonian there exists a $G$-invariant density matrix  $\rho_G$   which is a solution of the Schr\"odinger equation  \eqref{SSErho}.

\medskip

\noindent
{\it Proof.} Consider a (not necessarily $G$-invariant) density matrix $\rho$ which is a solution of eq. \eqref{SSErho} corresponding to a given $E$. A $G$-invariant  solution $\rho_G$ can be obtained from $\rho$ by averaging over the group $G$ with the Haar measure $d\mu(U)$~\cite{Fulton1991,Bartlett2003}:
\begin{equation}\label{averaged}
\rho_G=\int\limits_G U\rho\, U^\dag d\mu(U)
\end{equation}
where the normalization condition $\int_G 1\,d\mu(U)=1$ is implied. It is easy to see that thus obtained $\rho_G$  is indeed a legitimate density matrix (i.e. satisfies conditions \eqref{rho conditions}) and is invariant under the group $G$ ( i.e.  $U\,\rho_G\,U^\dagger =\rho_G~~~~~\forall ~~U\in G$).~\qed

\medskip

Obtaining $G$-invariant objects by  averaging over the group $G$ with the Haar measure   is a standard tool of the group theory \cite{Fulton1991}, and mixed states of the form  \eqref{averaged} naturally appear in various resource theories \cite{Bartlett2003,Vaccaro2008,Gour2009,Hall2012}.  It should be emphasized, however,  that we use such averaging only as a formal tool to prove the existence result contained in the above Lemma. In practice, we propose to ensure  the $G$-invariance by expanding the density matrix in $G$-invariant basis operators, without explicitly performing the averaging \eqref{averaged}. We apply this approach to specific examples in what follows.


It is worth highlighting why the averaging over the group analogous to that in eq. \eqref{averaged} can not be applied directly to vectors in the Hilbert space. This is because such averaging does not conserve the normalization, and one can obtain a zero vector (which lacks physical interpretation) as a result. This indeed happens, as can be seen in the trivial example of a single spin $1/2$ with $H=0$ discussed above. In contrast, averaging \eqref{averaged} of density matrices conserves the trace and thus the normalization.

The Schr\"odinger equation \eqref{SSErho} entails 
\be\label{anti-Hermitian}
[H,\rho]=0.
\ee
In fact, this equation holds not only for the solutions of the SSE$_{\rho}$ but for any stationary state, i.e. a state not evolving under the Liouville–von Neumann equation. Eq. \eqref{anti-Hermitian} is widely used to obtain constraints on expectation values of various observables in equilibrium \cite{Hirschfelder,Mukherjee,Mazziotti}. For our purposes it is essential that this equation does not contain $E$ and can be used to reduce the dimensionality of the more computationally demanding eigenvalue problem. This is discussed in more detail in the next section.


In the rest of the paper we illustrate the concept of the SSE$_\rho$  by considering specific spin systems.

\section{System of spins $1/2$ with the Heisenberg interaction\label{sec:Heisenberg}}

In the present section we specialise the SSE$_{\rho}$ for a system of $N$ spins with the Heisenberg interaction. The Hamiltonian of this system reads
\begin{equation}
	H = \sum_{i<j}^{} J_{ij} \left( {{{\bm \sigma}_i}{{\bm \sigma}_j}} \right) ,~~~~~i,j=1,2,...,N,
\end{equation}
where ${\bm \sigma}_i$ is the vector consisting of three Pauli matrices of the $i$'th spin,  $ J_{ij} $ is the coupling constant between $i$'th and $j$'th spins and $\left( {{{\bm \sigma}_i}{{\bm \sigma}_j}} \right)$ is the corresponding scalar product of sigma-matrices. This Hamiltonian is invariant with respect to a global $SU(2)$ symmetry, in other words, to the simultaneous rotations of all spins. In addition, it is $T$-invariant, i.e. invariant with respect to the inversion of time. This Hamiltonian, apart from being a popular theoretical playground, is of practical importance in material science, both for finite \cite{schnalle2010calculating} and for infinite~\cite{manousakis1991spin}~$N$.

Due to the presence of the above symmetries, we can look for a solution $\rho$ of the $SSE_\rho$ which is constructed of scalar products of sigma matrices. To this end, we define a multi-index ${\cal A}$ enumerating the set of pairs $(i_p,j_p)$:
\be\label{multi-index}
{\cal A}=\left( {{i_1},{j_1}} \right)...\left( {{i_m},{j_m}} \right),~~~1\leq m\leq [N/2],
\ee
where $[N/2]$ is the integer part of $N/2$, while $i_p$ and $j_p$ enumerate spins and for any $p$ satisfy
\be\label{condiditons on indices}
i_p<i_{p+1}, ~~~~j_p>i_p, ~~~~ j_p \ne i_l, j_l~~~\forall~l \ne p,~~~~1\leq i_p,\,j_p\leq N.
\ee
These conditions ensure that the sum over ${\cal A}$ runs over all distinct sets of pairs of indices in which each index is found at most once. We denote the number of pairs in  ${\cal A}$ by  $|{\cal A}|$ (e.g. $|{\cal A}|=m$ in eq. \eqref{multi-index}). Finally, we define an operator $A_{\cal A}$ as a product of $|{\cal A}|$ scalar products of Pauli matrices according to
\be\label{A}
A_{\cal A}=\left( {{{\bm \sigma} _{{i_1}}}{{\bm \sigma} _{{j_1}}}} \right)\,\left( {{{\bm \sigma} _{{i_2}}}{{\bm \sigma} _{{j_2}}}} \right)\, ...\,
\left( {{{\bm \sigma} _{{i_m}}}{{\bm \sigma} _{{j_m}}}} \right),
\ee
where ${\cal A}$ is given by eq. \eqref{multi-index}. We supplement this definition by a convention $A_0\equiv1$.

Our ansatz for $\rho$ can now be written as
\be\label{ansatz}
\rho=\frac{1}{2^N}\left(a_0 A_0+ \sum_{\cal A} a_{\cal A} A_{\cal A}\right).
\ee
Here $a_{\cal A}$ are numerical coefficients and the sum is over all sets ${\cal A}$ of the form \eqref{multi-index},\eqref{condiditons on indices}. Note that normalization implies $ a_0 = 1 $. Obviously, such $\rho$ is both $SU(2)$-invariant and $T$-invariant. In fact, any $SU(2)$- and $T$-invariant operator with a unit trace can be represented in this form. Let us briefly explain why.  First, observe that due to the equalities
\begin{align}
{({{\bm{\sigma }}_1}{{\ssigma}_2})^2} =& \;\; 3 - 2({{\ssigma}_1}{{\ssigma}_2}),\nonumber \\
({{\ssigma}_1}{{\ssigma}_2})({{\ssigma}_2}{{\ssigma}_3}) = &\;\;\;({{\ssigma}_1}{{\ssigma}_3}) - i({{\ssigma}_1}{{\ssigma}_2}{{\ssigma}_3}),\label{algebraic relations}
\end{align}
one can avoid terms with repeating spin indexes, as is indeed the case in eq. \eqref{ansatz}. Here $({{\ssigma}_1}{{\ssigma}_2}{{\ssigma}_3})$ is the mixed product of vectors consisting of Pauli matrices of three spins. Further, observe that the mixed product changes its sign under time inversion and thus does not enter $\rho$. As for the even powers of mixed products, they can always be expressed through the scalar products \cite{Uskov}. 

One can now substitute the ansatz \eqref{ansatz} into the stationary Schr\"odinger equation \eqref{SSErho}. Exploiting formulae \eqref{algebraic relations}, after straightforward but tedious calculations  one obtains the following equations for the coefficients $a_\A$:

\begin{align}\label{equation Heisenberg}
 E\,a_{0}&  = 3 \sum_{i<j}{ J_{ij} a_{(i,j)} },  \\
 \label{equation Heisenberg 2}
 E\,a_{\A}&  =
 \sum_{l=1}^{|\A|-1}\sum_{m=l+1}^{|\A|} \left(
 (J_{i_m j_l}+J_{i_l j_m}-J_{i_l j_l}-J_{i_m j_m})a_{(i_l,i_m)(j_l,j_m)\A^{\overline{l,m}}} \,\,
 \right. \\ \nonumber
 &+\left. (J_{i_l i_m}+J_{j_l j_m}-J_{i_l j_l}-J_{i_m j_m})a_{(i_l,j_m)(j_l,i_m)\A^{\overline{l,m}}}  \right)  \\\nonumber
 & +~~\sum_{\mathclap{\substack{ p,\,q:\, p<q \\\nonumber \, p,q\notin \A }}}
 \,\, \left(
 3\,J_{pq}\, a_{(p,q)\A}
 + \sum_{m=1}^{|\A|} J_{pq} (a_{(i_m,p)(q,j_m)\A^{\overline{m}}}+a_{(q,i_m)(j_m,p)\A^{\overline{m}}})
 \right) \nonumber\\
  & + \sum_{p\notin \A} \,\, \sum_{m=1}^{|\A|}  \left(J_{p j_m} a_{(p,i_m) \A^{\overline m}} + J_{p i_m} a_{(p,j_m) \A^{\overline m}} \right) +
 \sum_{m=1}^{|\A|} {J_{i_m j_m} \left( a_{\A^{\overline m}}-2a_{\A} \right)}, \nonumber
 \end{align}

\begin{align}\label{anti-Hermitian Heisenberg}
0 &=    \sum\limits_{p  \notin \A}^{} {\left[ {({J_{pj}} - {J_{pi}}){a_{\left( {ij} \right)\left( {pk} \right)\A }} + ({J_{pk}} - {J_{pj}}){a_{\left( {ip} \right)\left( {jk} \right)\A^{ }}} + ({J_{pi}} - {J_{pk}}){a_{\left( {ik} \right)\left( {pj} \right)\A^{ }}}} \right]} \\
&-  \left( {({J_{ik}} - {J_{jk}}){a_{\left( {ij} \right)\A^{ }}} + ({J_{jk}} - {J_{ij}}){a_{\left( {ik} \right)\A^{ }}} + ({J_{ij}} - {J_{ik}}){a_{\left( {jk} \right)\A^{ }}}} \right),
~~{\forall~i,j,k} \notin \A ~{ \rm and } ~i<j<k.  \nonumber
\end{align}
The multi-index $(i_l,j_m)(j_l,i_m)\A^{\overline{l,m}}$ is obtained from $\A$ by dropping pairs $(i_l,j_l)$,  $(i_m,j_m)$ and adding pairs $(i_l,j_m)$, $(j_l,i_m)$.
Other multi-indices used in the above equations are obtained from $\A$ analogously. 

Note that eq. \eqref{anti-Hermitian Heisenberg} is due to eq. \eqref{anti-Hermitian}. While eqs. \eqref{equation Heisenberg}, \eqref{equation Heisenberg 2} represent an eigenvalue problem, eq. \eqref{anti-Hermitian Heisenberg} is a homogeneous linear equation. Solving the latter is computationally less demanding than solving the former. Thus it can be numerically efficient to first eliminate as many variables as possible with the help of eq. \eqref{anti-Hermitian Heisenberg}, and then solve an eigenvalue problem with a smaller number of variables.

We remark that in the paradigm of the Contracted Schr\"odinger equation \cite{herbert2002extensivity} the spin Hamiltonian should be first turned into a fermionic Hamiltonian as in ref. \cite{schwerdtfeger2009convex-set}. While this is easily done for one-dimensional spin chains by means of the Jordan-Wigner transformation, in higher dimensions this leads to nonlocal interactions. In contrast, in our procedure transformation to fermionic representation is not required.

For illustrative purposes we apply the SSE$_\rho$ to the system of three spins:
\begin{align}
H &= {J_{12}}\left( {{\ssigma _1}{\ssigma _2}} \right) + {J_{23}}\left( {{\ssigma _2}{\ssigma _3}} \right) + {J_{13}}\left( {{\ssigma _1}{\ssigma _3}} \right) \label{H 3 spins}, \\
8\rho  &= a_0 + {a_{12}}\left( {{\ssigma _1}{\ssigma _2}} \right) + {a_{23}}\left( {{\ssigma _2}{\ssigma _3}} \right) + {a_{13}}\left( {{\ssigma _1}{\ssigma _3}} \right).
\end{align}
In this case eqs. \eqref{equation Heisenberg}, \eqref{anti-Hermitian Heisenberg} read
\be\label{eigenproblem 3 spins}
\begin{pmatrix}
0 & 3 J_{12} & 3 J_{23} & 3 J_{13}\\
J_{12} & -2 J_{12} &  J_{23} &  J_{13}\\
J_{23} & J_{12} & -2 J_{23} &  J_{13}\\
J_{13} & J_{12} &  J_{23} &  -2 J_{13}
\end{pmatrix}
\begin{pmatrix}
a_0 \\ a_{12} \\  a_{23} \\  a_{13}
\end{pmatrix}
=E
\begin{pmatrix}
a_0 \\ a_{12} \\  a_{23} \\  a_{13}
\end{pmatrix},
\ee
\begin{align}\label{antihermitian 3 spins}
0 &= {a_{12}}\left( {{J_{23}} - {J_{13}}} \right) + {a_{23}}\left( {{J_{13}} - {J_{12}}} \right) + {a_{13}}\left( {{J_{12}} - {J_{23}}} \right).
\end{align}
Note that the homogeneous linear equation  \eqref{antihermitian 3 spins} is, in principle, redundant, but in practice can be useful for simplifying the eigenproblem \eqref{eigenproblem 3 spins}, as discussed above. We also remark that the size of eigenproblem is twice smaller than what one would obtain by a straightforward application of the conventional Schr\"odinger equation \eqref{SSE} to the system of three spins $1/2$. This size is even more reduced if the Hamiltonian posses additional symmetries, see below. This can prove useful for exact diagonalization studies of small spin clusters, which can be of interest for understanding magnetic response of correlated materials \cite{lychkovskiy2018spin} (for alternative ways of  accounting for symmetries see \cite{schnalle2010calculating}).

\begin{figure}[t]
\begin{center}
 	\includegraphics[width= 0.5 \linewidth]{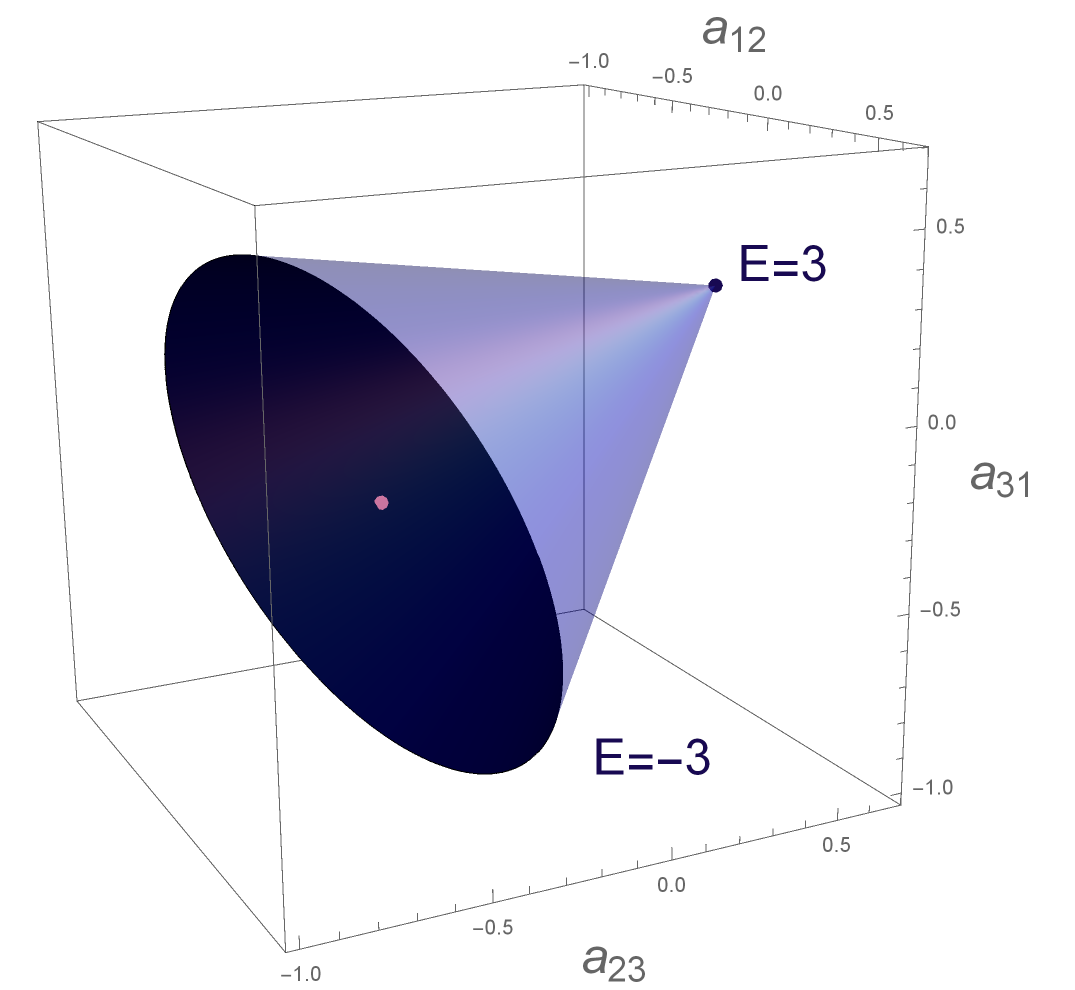}
\end{center}
\caption{\label{fig}
	The set of $SU(2)$- and $T$-invariant density matrices of three spins $1/2$ \cite{ilin2018squaring}. The solutions \eqref{E=3} and \eqref{E=-3} of the  SSE$_\rho$ with the Hamiltonian \eqref{H 3 spins} with $ {J_{12}} = {J_{23}} = {J_{13}} = 1 $ correspond respectively to the tip and to the base of the cone. The point in the center of the base of the cone corresponds to the maximally symmetric (permutation-invariant) solution of the form \eqref{E=-3} with $a_{12}=a_{23}=a_{13}=-1/3$.
}
\end{figure}

If $ {J_{12}} = {J_{23}} = {J_{13}} = 1 $, eq. \eqref{eigenproblem 3 spins} leads to two sets of solutions:
\begin{align}
	E &= 3,~~~~~~a_0=1,~~{\rm{      }}{a_{12}} = {a_{23}} = {a_{13}} = \frac{1}{3} \label{E=3}; \\
	E &~=  - 3,~~~a_0=1,~~{\rm{      }}{a_{12}} + {a_{23}} + {a_{13}} =  - 1 ,\label{E=-3}
\end{align}
see figure \ref{fig} for illustration.

\section{Total spin of $N$ spins $1/2$: Projections on invariant subspaces\label{sec:total spin}}
Total spin operator is formally equivalent to the Heisenberg Hamiltonian with long-range interactions.
We seek to solve the eigenproblem for total spin of N qubits
\begin{equation}\label{Ssquared}
	{\bm S^2}\rho  = \lambda \rho,
\end{equation}
with an additional constraint that $\rho$ is invariant under permutations of spins,
\begin{equation}\label{rho for Ssquared}
	\rho  = \frac{1}{{{2^N}}}\left(1+ \sum_{m=1}^{[N/2]}{a_m A_m}\right),
\end{equation}
where $A_m$ is the sum of all possible products~\eqref{A} of $m$ scalar products of sigma matrices. It can be easily found that $ {\bm S^2}=\frac{1}{4} (3N + 2{A_1}) $ and
\begin{align}
	{A_1}{A_m} &= (N - 2m + 2)(N - 2m + 1)\left( {\frac{3}{2} + (m - 1)\theta (N - 2m+1)} \right){A_{m - 1}} ~~~~~~~~~~~~~~~~~~~\\
	&+ 2m\left( {(N - 2m)\theta (N - 2m ) - 1} \right){A_m}
	\begin{array}{c}
	+ (m + 1)\theta (N - 2m - 1){A_{m + 1}},
	\end{array} \nonumber	m = 1,...[N/2].
\end{align}
 Here $\theta( x)$ is 1 if $ x>0 $ and 0 otherwise.
The above relation defines a tridiagonal matrix which has eigenvalues $\lambda = S (S+1)$, as we verified numerically. They lead to the recursive formula for the coefficients $a_m$:
\begin{align}
	{a_1} &= \frac{{4\lambda  - 3N}}{{3N(N - 1)}},\nonumber \\
	{a_2} &= \frac{{(4\lambda  - 7N + 12){a_1} - 2}}{{5(N - 2)(N - 3)}}\theta (N - 3),\nonumber \\
	{a_m} &= \frac{{(4\lambda  - (4m - 1)N + 8{m^2} - 12m + 4){a_{m - 1}} - 2(m - 1){a_{m - 2}}}}{{(N - 2m + 2)(N - 2m + 1)(2m + 1)}},~~~m=3,4,...[N/2].
\end{align}
It can be verified that thus obtained density matrices \eqref{rho for Ssquared} coincide up to normalization with the projections to invariant subspaces of ${\bm S^2}$.

\section{Systems of spins $1$ \label{sec:spin one}}

In this section we briefly outline how SSE$_\rho$ can be applied to systems of spins $1$. While for a spin  $1/2$  three  operators of spin projections along with the identity operator span the whole space of self-adjoint operators, this is not the case for a spin $1$. As a result, the ansatz for an invariant density matrix  becomes more complicated. Let us start from a system of two spins $1$ with the Hamiltonian invariant under rotations,
\be
H=(\Sb_1 \Sb_2),
\ee
where $\Sb_i$ is the spin at $i$'th site, $(\Sb_i \Sb_i)=2$. A general form of the rotationally-invariant density matrix reads
\begin{align}
\rho = a_0 + a_1 \left( {{{\bm S}_1}{{\bm S}_2}} \right) + a_2 \left( {{{\bm S}_1}{{\bm S}_2}} \right)^2.
\end{align}
Higher powers of the scalar product $(\Sb_1\Sb_2)$ are linearly dependent on the first two powers according to
\begin{align}
\left( {{{\bm S}_i}{{\bm S}_j}} \right)^3 = 2 +\left( {{{\bm S}_i}{{\bm S}_j}} \right) - 2 \left( {{{\bm S}_i}{{\bm S}_j}} \right)^2.
\end{align}
The $SSE_{\rho}$ then reads
\begin{align}\label{two spins1}
\left(
\begin{array}{ccc}
	0 & 0 &  2 \\
	1 & 0 &  1 \\
	0 & 1 & -2
\end{array}
\right)
\begin{pmatrix}
a_0 \\ a_1 \\ a_2
\end{pmatrix}
=E
\begin{pmatrix}
a_0 \\ a_1 \\ a_2
\end{pmatrix}
\end{align}
with the solution
\begin{align}
	E &= 1,& (a_0,a_1,a_2) &=(1/15,1/10,1/30),  \nonumber \\
	E &~=  - 1, & (a_0,a_1,a_2)&=(1/3,-1/6,-1/6),\nonumber \\
	E &~=  - 2,  &  (a_0,a_1,a_2)& =(-1/3,0,1/3) ,
\end{align}
where the normalization condition $\tr \rho=1$ is taken into account.

Now we turn to a case of three spins with interactions invariant under rotations,  time reversal and permutations. A density matrix respecting these symmetries can be parameterized as
\begin{align}\label{ansatz spins 1}
\rho = a_0
+ a_1 \sum_{i<j}
\left( {{{\bm S}_i}{{\bm S}_j}} \right)
+ a_{21} \sum_{i\neq j\neq k} \left( {{{\bm S}_i}{{\bm S}_j}}\right) \left( {{{\bm S}_j}{{\bm S}_k}}\right)
+ a_{22} \sum_{i<j} \left( {{{\bm S}_i}{{\bm S}_j}} \right)^2
+ a_3 \sum_{i\neq j\neq k} \left( {{{\bm S}_i}{{\bm S}_j}} \right) \left( {{{\bm S}_j}{{\bm S}_k}} \right)\left( {{{\bm S}_i}{{\bm S}_k}} \right),
\end{align}
where $ \sum\limits_{i\neq j\neq k} $ is the sum over six triples of distinct $i,j$ and $k$. All other polynomials constructed of scalar products are linear dependent on those presented in eq. \eqref{ansatz spins 1}. The normalization condition implies
\be\label{normalization spins 1}
27a_0+108 a_{22}+144 a_3 =1.
\ee
Thus the ansatz contains only four real  parameters (say, $a_1,a_{21},a_{22}, a_3$), to be compared to  $2\times 3^3-1=53$ real parameters required to parameterize a pure state in the Hilbert space of three spins $1$ without account for symmetries.

We consider two different three-spin Hamiltonians respecting the above symmetries. The first one is
\be
H=(\Sb_1 \Sb_2)+(\Sb_2 \Sb_3)+(\Sb_3 \Sb_1).
\ee
The SSE$_{\rho}$ for this Hamiltonian reads
\begin{align}\label{SSErho spins 1}
\left(
\begin{array}{ccccc}
	0 & 0 &  0 & 6 & 0 \\
	1 & 0 &  4  & 3 & 4 \\
	0 & 1 & 0 & 0 & 2 \\
	0 & 1 & 0 & -2 & 2 \\
    0 & 0 & 1 & 0 & -2
\end{array}
\right)
\begin{pmatrix}
a_0 \\ a_1 \\ a_{21}\\a_{22}\\a_{3}
\end{pmatrix}
=E
\begin{pmatrix}
a_0 \\ a_1 \\ a_{21}\\a_{22}\\a_{3}
\end{pmatrix}
\end{align}
The eigenvalues are $(-3,-2,0,3)$. The coefficient  $(a_0,a_1,a_{21},a_{22}, a_3)$ are also easily found from eqs. \eqref{SSErho spins 1} and  \eqref{normalization spins 1}, we omit them for brevity.

Another Hamiltonian we consider reads
\be
H=(\Sb_1 \Sb_2)^2+(\Sb_2 \Sb_3)^2+(\Sb_3 \Sb_1)^2.
\ee
The SSE$_{\rho}$ for this Hamiltonian is given by
\begin{align}\label{three spins1}
\left(
\begin{array}{ccccc}
	0 & 6 &  24 & -36 & -72 \\
	0 & 3 &  -2  & 2 & 4 \\
	0 & 0 & 3 & 0 & 0 \\
	1 & -2 & -8 & 15 & 24 \\
    0 & 0 & 1 & -1 & 1
\end{array}
\right)
\begin{pmatrix}
a_0 \\ a_1 \\ a_{21}\\a_{22}\\a_{3}
\end{pmatrix}
=E
\begin{pmatrix}
a_0 \\ a_1 \\ a_{21}\\a_{22}\\a_{3}
\end{pmatrix}
\end{align}
The eigenvalues are $(3,5,8)$.

For an arbitrary number of spins $1$ an ansatz for states invariant under rotations, permutations and time reversal has a form analogous to eq. \eqref{ansatz spins 1}: It contains symmetric polynomials in scalar products of spin operators, each spin entering each term of this polynomial at most twice. If the system lacks the permutation symmetry, the polynomials need not be symmetric. This ansatz can be readily used to obtain a SSE$_\rho$ for any number of spins.


\section{Summary}

We have studied the properties and merits of the stationary Schr\"odinger equation \eqref{SSErho} with density matrices instead of wave functions. This equation produces the same spectrum of eigenvalues as the conventional Schr\"odinger equation. The main advantage of eq.  \eqref{SSErho} shows up when the Hamiltonian is invariant under some symmetry group which induces degeneracies of the spectrum. In this case for any eigenenergy one can choose a solution of eq.  \eqref{SSErho} which is invariant under the symmetry group. This is in contrast to the conventional Schr\"odinger equation, where the spontaneous symmetry breaking can prevent one from finding an invariant eigenvector. We have exemplified eq.  \eqref{SSErho}  by applying it to a system of spins $1/2$ with the Heisenberg interactions on an arbitrary lattice. Further, we have applied an equation analogous to  eq.  \eqref{SSErho} to find invariant subspaces of the operator of the total spin of $N$ spins $1/2$. Finally, we outlined how the same technique can be applied to higher spins. We conclude by a remark that it can be interesting to extend the methods of the present paper to the time-dependent Schr\"odinger equation and to master equations describing evolution of open quantum systems. The latter topic is addressed in a spirit somewhat similar to that in the present paper in the recent article~\cite{liniov}.
\section*{Acknowledgements.} The work was supported by the Russian Science Foundation under the grant No. 17-71-20158.

\end{document}